\documentclass[12pt]{article}
\usepackage{graphicx}
\usepackage{lscape}
\usepackage{citesort}
\usepackage{amssymb}
\usepackage{appendix}
\usepackage{multirow}

\usepackage{mathrsfs}
\begin{document}
\def\qq{\langle \bar q q \rangle}
\def\uu{\langle \bar u u \rangle}
\def\dd{\langle \bar d d \rangle}
\def\sp{\langle \bar s s \rangle}
\def\GG{\langle g_s^2 G^2 \rangle}
\def\Tr{\mbox{Tr}}
\def\figt#1#2#3{
        \begin{figure}
        $\left. \right.$
        \vspace*{-2cm}
        \begin{center}
        \includegraphics[width=10cm]{#1}
        \end{center}
        \vspace*{-0.2cm}
        \caption{#3}
        \label{#2}
        \end{figure}
    }

\def\figb#1#2#3{
        \begin{figure}
        $\left. \right.$
        \vspace*{-1cm}
        \begin{center}
        \includegraphics[width=10cm]{#1}
        \end{center}
        \vspace*{-0.2cm}
        \caption{#3}
        \label{#2}
        \end{figure}
                }

\def\ds{\displaystyle}
\def\beq{\begin{equation}}
\def\eeq{\end{equation}}
\def\bea{\begin{eqnarray}}
\def\eea{\end{eqnarray}}
\def\beeq{\begin{eqnarray}}
\def\eeeq{\end{eqnarray}}
\def\ve{\vert}
\def\vel{\left|}
\def\ver{\right|}
\def\nnb{\nonumber}
\def\ga{\left(}
\def\dr{\right)}
\def\aga{\left\{}
\def\adr{\right\}}
\def\lla{\left<}
\def\rra{\right>}
\def\rar{\rightarrow}
\def\lrar{\leftrightarrow}
\def\nnb{\nonumber}
\def\la{\langle}
\def\ra{\rangle}
\def\ba{\begin{array}}
\def\ea{\end{array}}
\def\tr{\mbox{Tr}}
\def\ssp{{\Sigma^{*+}}}
\def\sso{{\Sigma^{*0}}}
\def\ssm{{\Sigma^{*-}}}
\def\xis0{{\Xi^{*0}}}
\def\xism{{\Xi^{*-}}}
\def\qs{\la \bar s s \ra}
\def\qu{\la \bar u u \ra}
\def\qd{\la \bar d d \ra}
\def\qq{\la \bar q q \ra}
\def\gGgG{\la g^2 G^2 \ra}
\def\q{\gamma_5 \not\!q}
\def\x{\gamma_5 \not\!x}
\def\g5{\gamma_5}
\def\sb{S_Q^{cf}}
\def\sd{S_d^{be}}
\def\su{S_u^{ad}}
\def\sbp{{S}_Q^{'cf}}
\def\sdp{{S}_d^{'be}}
\def\sup{{S}_u^{'ad}}
\def\ssp{{S}_s^{'??}}

\def\sig{\sigma_{\mu \nu} \gamma_5 p^\mu q^\nu}
\def\fo{f_0(\frac{s_0}{M^2})}
\def\ffi{f_1(\frac{s_0}{M^2})}
\def\fii{f_2(\frac{s_0}{M^2})}
\def\O{{\cal O}}
\def\sl{{\Sigma^0 \Lambda}}
\def\es{\!\!\! &=& \!\!\!}
\def\ap{\!\!\! &\approx& \!\!\!}
\def\md{\!\!\!\! &\mid& \!\!\!\!}
\def\ar{&+& \!\!\!}
\def\ek{&-& \!\!\!}
\def\kek{\!\!\!&-& \!\!\!}
\def\cp{&\times& \!\!\!}
\def\se{\!\!\! &\simeq& \!\!\!}
\def\eqv{&\equiv& \!\!\!}
\def\kpm{&\pm& \!\!\!}
\def\kmp{&\mp& \!\!\!}
\def\mcdot{\!\cdot\!}
\def\erar{&\rightarrow&}
\def\olra{\stackrel{\leftrightarrow}}
\def\ola{\stackrel{\leftarrow}}
\def\ora{\stackrel{\rightarrow}}

\def\simlt{\stackrel{<}{{}_\sim}}
\def\simgt{\stackrel{>}{{}_\sim}}


\title{
         {\Large
                 {\bf
                     Investigation  of the $B_c\rightarrow \chi_{c2} l \overline{\nu} $ transition via  QCD sum rules
                 }
         }
      }

\author{\vspace{1cm}\\
{\small K. Azizi$^a$ \thanks {e-mail: kazizi@dogus.edu.tr}\,, Y.
Sarac$^b$
\thanks {e-mail: ysoymak@atilim.edu.tr}\,\,, H.
Sundu$^c$ \thanks {e-mail: hayriye.sundu@kocaeli.edu.tr}} \\
{\small $^a$  Department of physics, Do\u gu\c s university, Ac{\i}badem-Kad{\i}k\"oy, 34722 Istanbul, Turkey} \\
{\small $^b$ Department of electrical and electronics engineering,
Atilim university, 06836 Ankara, Turkey} \\
{\small $^c$ Department of physics, Kocaeli university, 41380 Izmit,
Turkey}}
\date{}

\begin{titlepage}
\maketitle
\thispagestyle{empty}

\begin{abstract}
We calculate the transition  form factors of the semileptonic
$B_c\rightarrow \chi_{c2} l \overline{\nu} $ in the framework of 
 QCD sum rules taking into account the two-gluon condensate corrections. Using  the obtained results of form factors we estimate the decay widths and branching ratios related to this transition at all 
lepton channels. A comparison of the obtained results with the predictions of other non-perturbative approaches are also made. The orders of branching ratios for different lepton channels indicate that
the $B_c\rightarrow \chi_{c2} l \overline{\nu} $ transition can be studied at LHC using the collected or future data.

\end{abstract}

~~~PACS number(s): 11.55.Hx, 13.20.-v, 13.20.He
\end{titlepage}

\section{Introduction}

The heavy-light systems such as   $B_c$ mesons are promising frameworks to study  the perturbative and non-perturbative aspects of
QCD. Among the $B_c$ systems, the mass and lifetime of the pseudoscalar ground state $B_c$ meson have been measured via different experimental groups \cite{Abe,Aaltonen,Abazov} and a more precise
 measurement of these quantities  is now available in particle data group (PDG) \cite{JBeringer}.
Other possible  $B_c$ states (the scalar, vector, axial-vector and tensor) have not been observed yet, but they are expected to be produced at the Large Hadron Collider (LHC) in near future. 
 The  $B_c$ meson as a doubly heavy quark-antiquark bound state with explicit flavors constitutes a rich laboratory for examining the QCD potential
models and better understanding the weak decay mechanisms of the  heavy flavor hadrons. This  flavor asymmetric ground state ($b \bar c$)   is in the focus of much
attention   compared to the flavor-neutral heavy quarkonia states since  it only decays
via weak interactions. These properties provide a fertile  ground for this meson to be in agenda of different experiments.  It is expected that the LHC and super-B experiments will provide more data regarding 
 the pseudoscalar $B_c$ meson decays. This is a motivation for theoreticians to complete their studies on the decay 
channels of this meson.



One of the possible decay channels of the $B_c$ meson is its semi-leptonic transition to the charmonium $\chi_{c2}$ tensor meson which is expected to have a considerable contribution to the total decay width.
Our goal in this article is to study this decay channel and calculate some related  physical quantities. By applying the QCD sum rules as one of the applicable and attractive non-perturbative approaches,  we calculate
the transition form factors responsible for the semileptonic $B_c\rightarrow \chi_{c2} l \nu $ transition. In the calculations, we consider the two gluon condensate contributions and extend the previous theoretical
calculations on these contributions to include the tensor state for the first time. The interpolating current of the $\chi_{c2}$ tensor meson with quantum numbers $I^G(J^{PC})=0^+(2^{++})$ includes covariant derivatives
with respect to position (for more information about the properties of this meson see \cite{AlievKazim}), hence we start our calculations in the coordinate space then we transform the calculations to momentum space performing Fourier integrals. To suppress the contributions of the higher states and continuum,
we apply both  Borel transformation and continuum subtraction as  necessities  of the method. We use the  transition form factors, then, to estimate the decay widths and branching ratios of the transition under consideration
for different lepton channels. Note that this transition  has been previously studied via different approaches like  covariant light-front quark model (CLFQM) \cite{Wang2}, 
generalized instantaneous approach (GIA) \cite{CHChang},
  relativistic constituent quark model (RCQM) \cite{MAIvanov,MAIvanov1} and non-relativistic constituent quark model (NRCQM) \cite{EHernandez}. For some other decay channels of the  $B_c$ meson  studied via various
approaches such as   light cone and three-point QCD sum rules, relativistic quark model, covariant light-front quark model, the renormalization group method and non-relativistic constituent
quark model see 
\cite{Aliev1,Aliev4,Azizi2,Azizi3,Azizi5,Azizi6,Colangelo,Kiselev,Huang,Ivanov,Ebert1,Ebert2,Faessler,Chang,Ivanov2,Hernandez,Becher,Aglietti,Schwanda,Chang2,Colangelo2}.


The article contains three sections. Next section includes the
details of calculations of the form factors for  $B_c\rightarrow \chi_{c2} l \overline{\nu} $ via QCD sum rules. Section 3 encompasses our  numerical analysis of the form factors  
and estimation of the decay width and branching ratio of the decay channel under consideration. This section contains also our concluding remarks.

\section{ QCD sum rules for transition form factors}
In this section  the details of  calculations for the form factors are  presented.  The
$B_c\rightarrow \chi_{c2} l \overline{\nu} $ decay channel  is based on the tree-level $b\rightarrow c$  transition, whose effective Hamiltonian  is of the form
\begin{eqnarray}\label{Eq1} {\cal H}_{eff} = \frac{G_F}{\sqrt2}
V_{bc} ~\bar c \gamma_\mu (1-\gamma_5) b \bar l \gamma^\mu
(1-\gamma_5) \nu,
\end{eqnarray}
where $G_F$ is the Fermi coupling constant and $V_{cb}$ is element of the  CKM matrix.
By sandwiching the effective Hamiltonian between the initial and final states we obtain the following matrix elements for the vector and axial-vector parts of the 
 transition current   $J_\mu^{tr}=\bar c \gamma_\mu(1-\gamma_5) b$, parametrized in terms of form factors:
\begin{eqnarray}\label{matrixel1a} \langle \chi_{c2}(p') \md  J_\mu^{tr,V} \mid B_{c}(p)   \rangle= h(q^2)\epsilon_{\mu\nu\alpha\beta}
\epsilon'^{\nu\lambda}P_{\lambda}P^{\alpha}q^{\beta}~,
\end{eqnarray}
\begin{eqnarray}\label{matrixel1b} \langle \chi_{c2}(p') \md  J_\mu^{tr,A} \mid B_{c}(p)   \rangle= -i \Big\{K(q^2)\epsilon'^{*}_{\mu\nu}P^{\nu} +
\epsilon'^{*}_{\alpha\beta}P^{\alpha}P^{\beta}[P_\mu b_{+}(q^2)+ q_{\mu}b_{-}(q^2)]\Big\}~,
\end{eqnarray}
where $h(q^2)$, $K(q^2)$, $b_{+}(q^2)$ and $b_{-}(q^2)$ are transition form factors, $\epsilon'_{\alpha\beta}$ is the polarization tensor of the
$\chi_{c2}$ meson, $P_\mu=(p+p')_\mu$ and $q_\mu=(p-p')_\mu$.

To calculate the form factors  as the main goal of the present paper via QCD sum rules, we start with the following three-point correlation function:
\begin{eqnarray}\label{T}
\Pi_{\mu\alpha\beta} = i^2\int d^{4}xe^{-ipx}\int d^{4}ye^{ip'y}\langle 0 \mid
{\cal T}\{ {J^{\chi_{c2}}_{{\alpha\beta}}}(y)J_\mu^{tr,V(A)}(0) J^{\dag B_c}(x)\}\mid
0\rangle~,
\end{eqnarray}
where, ${\cal T}$ is the time ordering product. To proceed we need the interpolating currents of the initial and final mesonic states. Considering all quantum numbers, their interpolating currents can be written as
\begin{eqnarray}\label{tensorcurrent}
J^{\chi_{c2}}_{{\alpha\beta}}(y)=\frac{i}{2}[\bar{c}(y)\gamma_{\alpha}\olra{\cal D}_{\beta}~(y)
c(y)+\bar c(y) \gamma_{\beta}  \olra{\cal D}_{\alpha}(y) c(y)],
\end{eqnarray}
\begin{eqnarray}\label{tensorcurrent}
J^{B_c}(x)=\bar{c}(x)\gamma_{5}b(x),
\end{eqnarray}
where $ \olra{\cal D}_{\mu}(y)$
is a two-side covariant derivative acting on the left and right,
simultaneously. It is  defined as
\begin{eqnarray}\label{derivative}
\olra{\cal D}_{\mu}(y)=\frac{1}{2}\left[\ora{\cal D}_{\mu}(y)-
\ola{\cal D}_{\mu}(y)\right],
\end{eqnarray}
with
\begin{eqnarray}\label{derivative2}
\overrightarrow{{\cal D}}_{\mu}(y)=\overrightarrow{\partial}_{\mu}(y)-i
\frac{g}{2}\lambda^aA^a_\mu(y),\nonumber\\
\overleftarrow{{\cal D}}_{\mu}(y)=\overleftarrow{\partial}_{\mu}(y)+
i\frac{g}{2}\lambda^aA^a_\mu(y).
\end{eqnarray}
Where $\lambda^a$ are the Gell-Mann matrices and
$A^a_\mu(y)$ are the external  gluon fields.

From the general philosophy of the QCD sum rules, in order to find the form factors, we need to calculate the aforesaid correlation function in two different ways.
Firstly, we calculate it in terms of  hadronic degrees of freedom called phenomenological or physical representation. 
Secondly, we calculate it in terms of QCD degrees of freedom in  deep Euclidean region via operator product expansion (OPE). This representation of the correlation function is called 
the QCD side. By equating the coefficients of the selected structures from both sides,  QCD sum rules for the  form factors are obtained. 
To suppress the contributions coming from the higher states and continuum Borel transformation
as well as quark-hadron duality assumption are applied.

By inserting  appropriate complete sets of hadronic states into the correlation function and by isolating the ground state contribution, we obtain
\begin{eqnarray} \label{phys1} \Pi_{\mu\alpha\beta}^{PHYS}&=&\frac{\langle 0 \mid
 J^{\chi_{c2}}_{{\alpha\beta}}(0)\mid
\chi_{c2}(p')\rangle \langle \chi_{c2}(p') \mid J_\mu^{tr,V(A)}\mid
B_c(p)\rangle \langle B_c(p)\mid J^{\dag}_{B_c}(0)\mid
0\rangle}{(p'^2-m_{\chi_{c2}}^2)(p^2-m_{B_c}^2)}+\cdots~,
\end{eqnarray}
where,  $\cdots$ represents the contributions of the higher states and continuum. To proceed we need to also define the matrix elements
\begin{eqnarray}\label{mat1}
\langle 0 \mid
 J^{\chi_{c2}}_{{\alpha\beta}}(0)\mid
\chi_{c2}(p')\rangle=f_{\chi_{c2}}
m_{\chi_{c2}}^3\epsilon'_{\alpha\beta},
\end{eqnarray}
and
\begin{eqnarray}\label{mat2}
\langle B_c (p)\mid
 J^\dag_{B_{c}}(0)\mid
0\rangle=-i
\frac{f_{B_{c}} m_{B_c}^2}{m_c+m_b},
\end{eqnarray}
where $f_{\chi_{c2}}$ and $f_{B_{c}}$ are leptonic decay constants of
$\chi_{c2}$ and $B_c$ mesons, respectively. Using all matrix elements given in Eqs.~(\ref{matrixel1a}), (\ref{matrixel1b}), (\ref{mat1})
and (\ref{mat2}) in Eq.(\ref{phys1}),
the final representation of the correlation function for the  physical side is obtained as
\begin{eqnarray}\label{phen2}
\Pi _{\mu\alpha\beta}^{PHYS}&=&\frac{f_{\chi_{c2}}f_{B_c}
m_{B_c}^2 m_{\chi_{c2}}}
{8(m_b+m_c)(p'^2-m_{\chi_{c2}}^2)(p^2-m_{B_c}^2)}
\Bigg\{\frac{2}{3}\Big[-\Delta K(q^2)+\Delta'b_{-}(q^2)
\Big]q_{\mu}g_{\beta\alpha}\nonumber \\
 &+&\frac{2}{3}\Big[(\Delta-4m_{\chi_{c2}}^2) K(q^2)+\Delta'b_{+}(q^2)
\Big]P_{\mu}g_{\beta\alpha}+i(\Delta-4m_{\chi_{c2}}^2)
h(q^2)\varepsilon_{\lambda\nu\beta\mu}P_{\lambda}P_{\alpha}q_{\nu}
\nonumber \\
&+&\Delta K(q^2)q_{\alpha}g_{\beta\mu}+ \mbox{other
structures}\Bigg\}+...,
\end{eqnarray}
where
\begin{eqnarray}\label{phen2}
\Delta &=&m_{B_c}^2+3m_{\chi_{c2}}^2-q^2,
\nonumber \\
\Delta'
&=&m_{B_c}^4-2m_{B_c}^2(m_{\chi_{c2}}^2+q^2)+(m_{\chi_{c2}}^2-q^2)^2,
\end{eqnarray}
and we have kept only the structures which we are going to select in order to find the corresponding form factors. Note that for obtaining the above representation of the physical side,
we have performed summation over the  polarization tensor  using
\begin{eqnarray}\label{polarizationt1}
\varepsilon_{\mu\nu}\varepsilon_{\alpha\beta}^*=\frac{1}{2}T_{\mu\alpha}T_{\nu\beta}+
\frac{1}{2}T_{\mu\beta}T_{\nu\alpha}
-\frac{1}{3}T_{\mu\nu}T_{\alpha\beta},
\end{eqnarray}
where
\begin{eqnarray}\label{polarizationt2}
T_{\mu\nu}=-g_{\mu\nu}+\frac{q_\mu q_\nu}{m_{\chi_{c2}}^2}.
\end{eqnarray}

In QCD side the  correlation function 
is calculated in deep Euclidean region where  $p^2\rightarrow -\infty$ and $p'^2\rightarrow -\infty$ via  OPE. Substituting the explicit form of
the interpolating currents into the correlation function and contracting out all quark pairs via Wick's theorem, we obtain
\begin{eqnarray}\label{correl.func.2}
\Pi^{QCD} _{\mu\alpha\beta}&=&\frac{i^3}{4}\int d^{4}x\int
d^{4}ye^{-ip\cdot x}e^{ip'\cdot y}
\nonumber \\
&\times& \Bigg\{Tr\left[S_c^{ca}(x-y)\gamma_\alpha\olra{\cal
D}_{\beta}(y)
S_c^{ab}(y)\gamma_\mu(1-\gamma_5)S_b^{bc}(-x)\gamma_5\right]+
\left[\beta\leftrightarrow\alpha\right]
\Bigg\},\nnb\\
\end{eqnarray}
where $ S_Q(x)$ with $Q=b$ or $c$ is the heavy quark propagator. It is given by
\cite{Reinders85}:
\begin{eqnarray}
S_{Q{ij}}(x)&=&\frac{i}{(2\pi)^4}\int d^4k e^{-ik \cdot x} \left\{
\frac{\delta_{ij}}{\!\not\!{k}-m_Q}
-\frac{g_sG^{\alpha\beta}_{ij}}{4}\frac{\sigma_{\alpha\beta}(\!\not\!{k}+m_Q)+
(\!\not\!{k}+m_Q)\sigma_{\alpha\beta}}{(k^2-m_Q^2)^2}\right.\nonumber\\
&&\left.+\frac{\pi^2}{3} \langle \frac{\alpha_sGG}{\pi}\rangle
\delta_{ij}m_Q \frac{k^2+m_Q\!\not\!{k}}{(k^2-m_Q^2)^4}
+\cdots\right\} \, .
\end{eqnarray}
 Although being very small compared to the perturbative part we include the contribution coming from the gluon
 condensate terms as non-perturbative effects. Replacing the explicit expression of the propagator in Eq.~(\ref{correl.func.2}) and performing  integrals
 over $x$ and $y$ (the details of calculations can be found in Appendix A),  we find the  QCD
 side as
 \begin{eqnarray}\label{QCDside}
\Pi^{QCD}_{\mu\alpha\beta}
&=&\Big(\Pi^{pert}_1(q^2)+\Pi^{non-pert}_1(q^2)\Big)q_{\alpha}g_{\beta\mu}+
\Big(\Pi^{pert}_2(q^2)+\Pi^{non-pert}_2(q^2)\Big)q_{\mu}g_{\beta\alpha}\nonumber \\
&+&
\Big(\Pi^{pert}_3(q^2)+\Pi^{non-pert}_3(q^2)\Big)P_{\mu}g_{\beta\alpha}+
\Big(\Pi^{pert}_4(q^2)+\Pi^{non-pert}_4(q^2)\Big)\varepsilon_{\lambda\nu\beta\mu}P_{\lambda}
P_{\alpha}q_{\nu}
\nonumber \\
&+&\mbox{other\,\,\, structures},
\end{eqnarray}
where $\Pi^{pert}_i(q^2)$ with $i=1,2,3,4$ are the perturbative parts of the coefficients of the selected structures. They are expressed  in terms of  double dispersion
integrals  as
\begin{eqnarray}\label{QCDside1}
\Pi^{pert}_i(q^2)=\int^{}_{}ds\int^{}_{}ds'
\frac{\rho_i(s,s',q^2)}{(s-p^2)(s'-p'^2)},
\end{eqnarray}
where the spectral densities $\rho_i(s,s',q^2)$ are obtained by taking the imaginary parts of the $\Pi^{pert}_i$
functions, i.e.,
$\rho_i(s,s',q^2)=\frac{1}{\pi}Im[\Pi^{pert}_i]$.
The spectral densities corresponding to  four different Dirac structures shown in Eq.~(\ref{QCDside})
 are obtained as 
\begin{eqnarray}\label{rho}
\rho_1(s,s',q^2)&=& \int_{0}^{1}dx
\int_{0}^{1-x}dy~~\frac{9\Big[m_b(4x+2y-1)+m_c(8x+4y-3)\Big]}{128\pi^2},
\nonumber \\
\rho_2(s,s',q^2)&=& \int_{0}^{1}dx
\int_{0}^{1-x}dy~~\frac{9\Big[-m_b(4x+2y-1)+m_c(4x+2y-3)\Big]}{64\pi^2},
\nonumber \\
\rho_3(s,s',q^2)&=& \int_{0}^{1}dx
\int_{0}^{1-x}dy~~\frac{9\Big[m_b(2y-1)-m_c(1+2y)\Big]}{64\pi^2},
\nonumber \\
\rho_4(s,s',q^2)&=&0.
\end{eqnarray}

The QCD sum rules for  form factors are attained by matching the coefficients of the same structures from both
sides of the correlation function. After applying  double Borel transformation with respect to the initial and final momentum squared as well as continuum subtraction, we get
 the following sum rules for form factors:
\begin{eqnarray}\label{K}
K(q^2)&=&\frac{8(m_b+m_c)}{f_{B_c}
f_{\chi_{c2}}m_{B_c}m_{\chi_{c2}}^2(m_{B_c}^2+3m_{\chi_{c2}}^2-q^2)}e^{\frac{m_{B_c}^2}{M^2}}
e^{\frac{m_{\chi_{c2}}^2}{M'^2}}
\nonumber \\
&&\Bigg\{\int^{s_0}_{(m_b+m_c)^2}ds
\int^{s_{0}^{'}}_{4m_c^2}ds'~~
e^{\frac{-s}{M^2}}e^{\frac{-s'}{M'^2}}\rho_1(s,s',q^2)\theta[L(s,s',q^2)]
+\widehat{\textbf{B}}\Pi_1^{non-pert}\Bigg\}
\nonumber \\
b_{-}(q^2)&=&\frac{12(m_b+m_c)}{f_{B_c}
f_{\chi_{c2}}m_{B_c}^2m_{\chi_{c2}}\Big(m_{B_c}^4+(m_{\chi_{c2}}^2-q^2)^2-2m_{B_c}^2
(m_{\chi_{c2}}^2+q^2)\Big)} e^{\frac{m_{B_c}^2}{M^2}}
e^{\frac{m_{\chi_{c2}}^2}{M'^2}}
\nonumber \\
&\times&\Bigg\{\int^{s_0}_{(m_b+m_c)^2}ds
\int^{s_{0}^{'}}_{4m_c^2}ds'
e^{\frac{-s}{M^2}}e^{\frac{-s'}{M'^2}}\rho_2(s,s',q^2)\theta[L(s,s',q^2)]
+\widehat{\textbf{B}}\Pi_2^{non-pert}
\nonumber \\
&+&e^{\frac{-m_{B_c}^2}{M^2}}
e^{\frac{-m_{\chi_{c2}}^2}{M'^2}}\frac{f_{B_c}
f_{\chi_{c2}}m_{B_c}^2m_{\chi_{c2}}(m_{B_c}^2+3m_{\chi_{c2}}^2-q^2)}{12(m_b+m_c)}K(q^2)
\Bigg\}
\nonumber \\
b_{+}(q^2)&=&\frac{12(m_b+m_c)}{f_{B_c}
f_{\chi_{c2}}m_{B_c}^2m_{\chi_{c2}}\Big(m_{B_c}^4+(m_{\chi_{c2}}^2-q^2)^2-2m_{B_c}^2
(m_{\chi_{c2}}^2+q^2)\Big)} e^{\frac{m_{B_c}^2}{M^2}}
e^{\frac{m_{\chi_{c2}}^2}{M'^2}}
\nonumber \\
&\times&\Bigg\{\int^{s_0}_{(m_b+m_c)^2}ds
\int^{s_{0}^{'}}_{4m_c^2}ds'
e^{\frac{-s}{M^2}}e^{\frac{-s'}{M'^2}}\rho_3(s,s',q^2)\theta[L(s,s',q^2)]
+\widehat{\textbf{B}}\Pi_3^{non-pert}
\nonumber \\
&+&e^{\frac{-m_{B_c}^2}{M^2}}
e^{\frac{-m_{\chi_{c2}}^2}{M'^2}}\frac{f_{B_c}
f_{\chi_{c2}}m_{B_c}^2m_{\chi_{c2}}(m_{\chi_{c2}}^2-m_{B_c}^2+q^2)}{12(m_b+m_c)}K(q^2)
\Bigg\}
\nonumber \\
h(q^2)&=&\frac{8(m_b+m_c)}{f_{B_c}
f_{\chi_{c2}}m_{B_c}^2m_{\chi_{c2}}\Big(m_{\chi_{c2}}^2-m_{B_c}^2+q^2\Big)}
e^{\frac{m_{B_c}^2}{M^2}} e^{\frac{m_{\chi_{c2}}^2}{M'^2}}
\nonumber \\
&\times& \Big\{\int^{s_0}_{(m_b+m_c)^2}ds
\int^{s_{0}^{'}}_{4m_c^2}ds'
e^{\frac{-s}{M^2}}e^{\frac{-s'}{M'^2}}\rho_4(s,s',q^2)\theta[L(s,s',q^2)]
+\widehat{\textbf{B}}\Pi_4^{non-pert} \Big\}
\end{eqnarray}
where $M^2$ and $M^{'2}$ are Borel mass parameters; and  $s_0$ and $s'_0$ are continuum thresholds in the initial and final channels, respectively. The function $L(s,s',q^2)$ is given by 
\begin{eqnarray}\label{L}
L(s,s',q^2)=s'x-s'x^2-m_c^2x-m_b^2y+sy+q^2xy-sxy-s'xy-sy^2.
\end{eqnarray}
The functions $\widehat{\textbf{B}}\Pi_i^{non-pert}$ are very lengthy functions and we present only the explicit expression of the $\widehat{\textbf{B}}\Pi_1^{non-pert}$ in Appendix B.

\section{Numerical Results}
In this section we  numerically  analyze  the QCD sum rules for
form factors obtained in the previous section and look for the  fit functions of the form factors in terms of $q^2$ in whole physical region, which are then used to estimate 
the decay width and branching ratio of the transition under consideration. For this aim we use the following values for some input  parameters: $
m_{\chi_{c2}}=(3556.20\pm0.09)$~ MeV, $ m_{B_c}
=(6.277\pm0.006)$~GeV \cite{JBeringer}, $ f_{B_c} =(476\pm27)$~MeV
\cite{Veliev}, $ f_{\chi_{c2}}=0.0111\pm0.0062 $ \cite{AlievKazim},
$ G_{F} = 1.17\times 10^{-5}$~GeV$^{-2}$, $
V_{cb}=(41.2\pm1.1)\times 10^{-3} $ and
 $ \tau_{B_{c}} =(45.3\pm4.1)\times10^{-14}$~s \cite{JBeringer}. 

In addition to the above input parameters, the sum rules for the form factors include  four auxiliary parameters: two Borel mass parameters $M^2$ and $M'^2$ as well as
two continuum thresholds $s_0$ and $s'_0$. Since these are not physical
parameters, the results of form factors should be practically independent of
them. Therefore their working regions are 
determined such that the results of the form factors depend weakly on these parameters. The continuum
thresholds are not totally capricious but they are related to the energy of the first excited state in  initial
and final mesonic channels. This  consideration in our case
leads to the intervals $43~\mbox{GeV$^2$}\leq
s_0\leq49~\mbox{GeV$^2$}$ and $15~\mbox{GeV$^2$}\leq
s'_0\leq17~\mbox{GeV$^2$}$ for the continuum thresholds.

\begin{figure}[h!]
\includegraphics[totalheight=6cm,width=7cm]{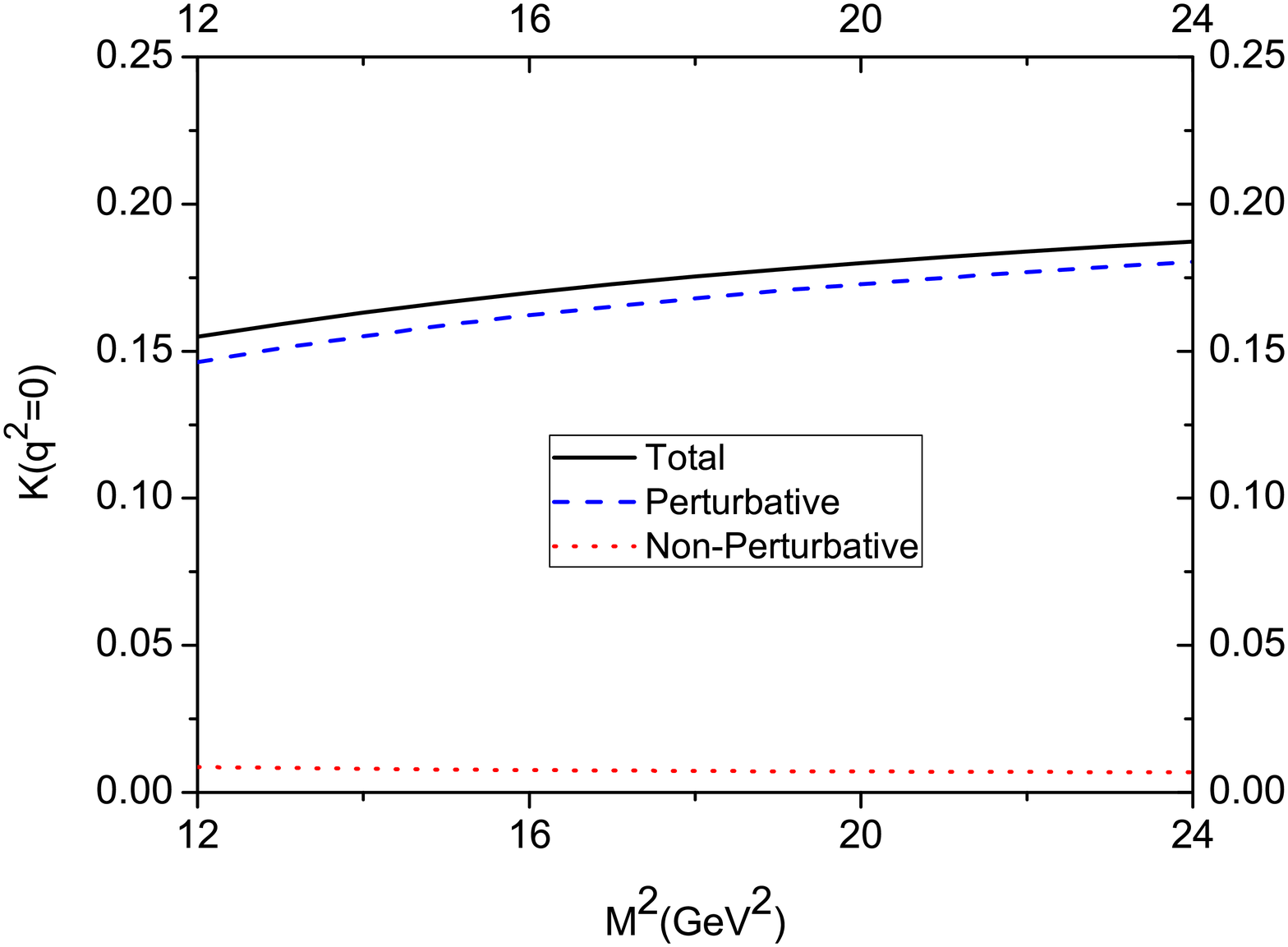}
\includegraphics[totalheight=6cm,width=7cm]{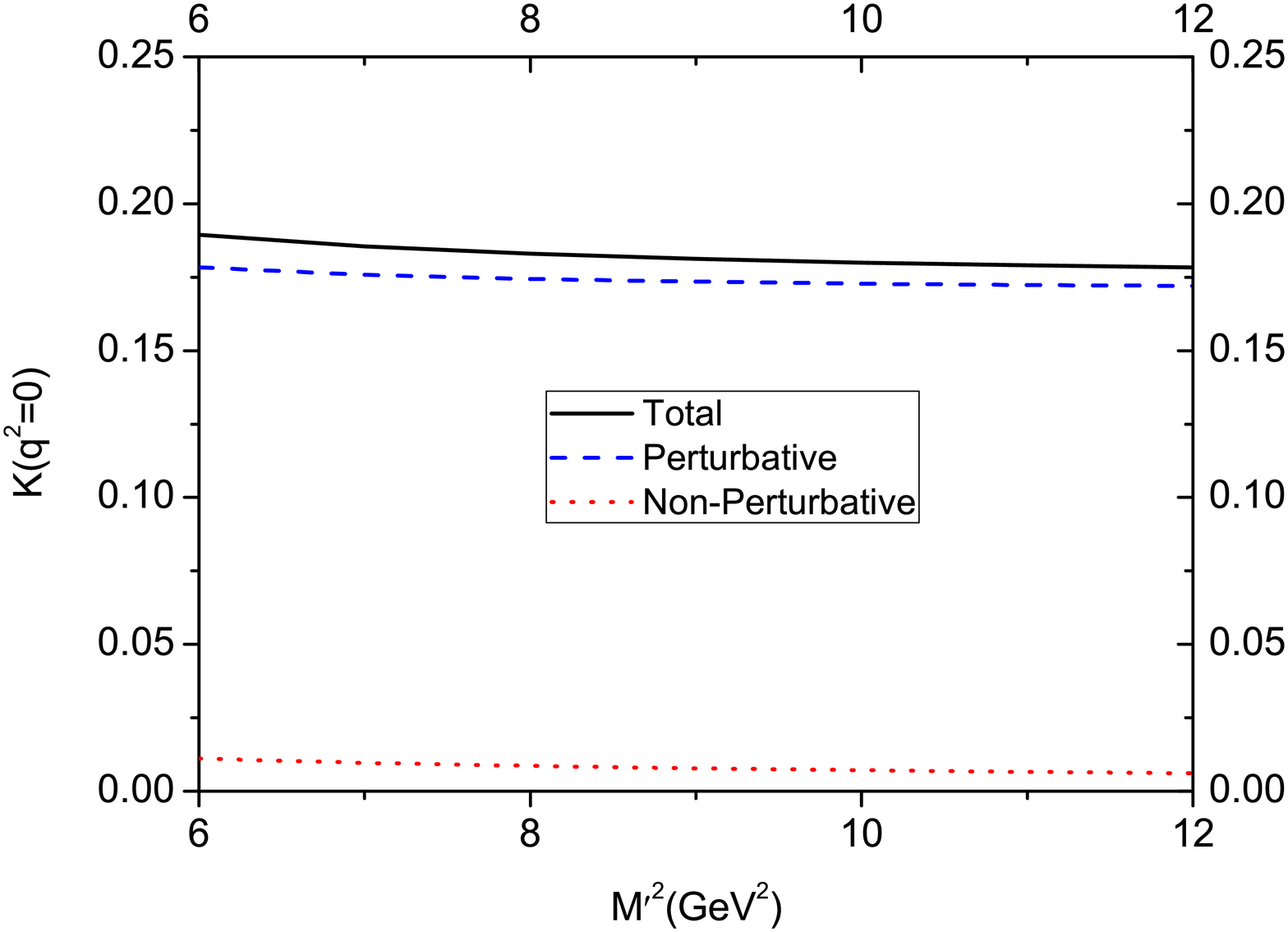}
\caption{\textbf{Left:} K($q^2=0$) as a function of the Borel mass
parameter $M^2$ at $M^{\prime^2}=10~GeV^2$. \textbf{Right:}
 K($q^2=0$) as a function of the
Borel mass parameter $M^{\prime^2}$ at $M^{2}=20~GeV^2$. } \label{KMsq}
\end{figure}

\begin{figure}[h!]
\begin{center}
\includegraphics[totalheight=8cm,width=10cm]{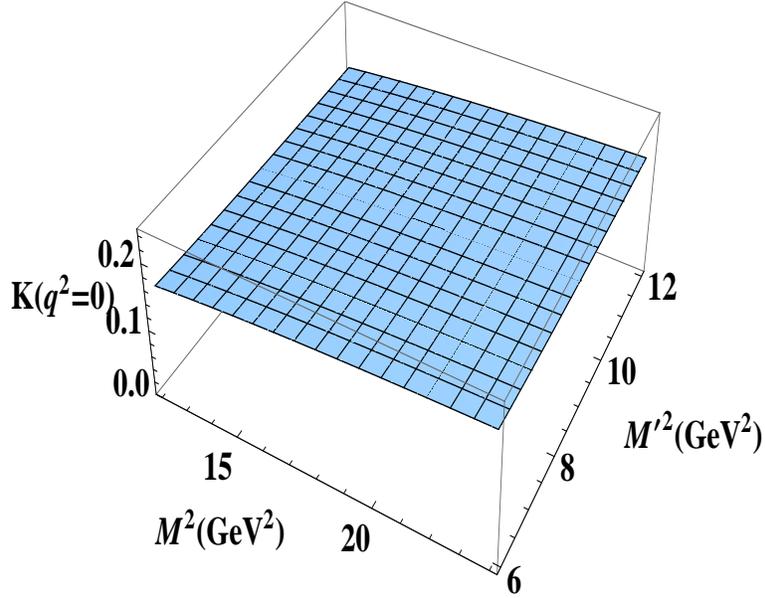}
\end{center}
\caption{ K($q^2=0$) as a function of the Borel mass
parameters $M^2$ and $M^{\prime^2}$. 
}
  \label{Kqsq1}
\end{figure}

 The working regions for the  Borel mass parameters are obtained by demanding that  the contributions of the higher states
and continuum are sufficiently suppressed and  the contributions
of the operators with higher mass dimensions are small compared to those having leading dimensions. These requirements lead to the intervals  $12~\mbox{GeV$^2$}\leq M^2\leq
24~\mbox{GeV$^2$}$ and $6~\mbox{GeV$^2$}\leq M'^2\leq
12~\mbox{GeV$^2$}$, for the Borel mass parameters in initial and final channels, respectively. In order to see how our results depend on the Borel parameters we present
 the dependence of the form factor $K(q^2)$, as an example, at $q^2=0$ on these parameters in figs.~\ref{KMsq} and \ref{Kqsq1}. From these figures we see that our results not only weakly depend on the Borel parameters,
 but also the perturbative contributions exceed the non-perturbative contributions and the series of form factors are convergent.

\begin{figure}[h!]
\begin{center}
\includegraphics[totalheight=8cm,width=10cm]{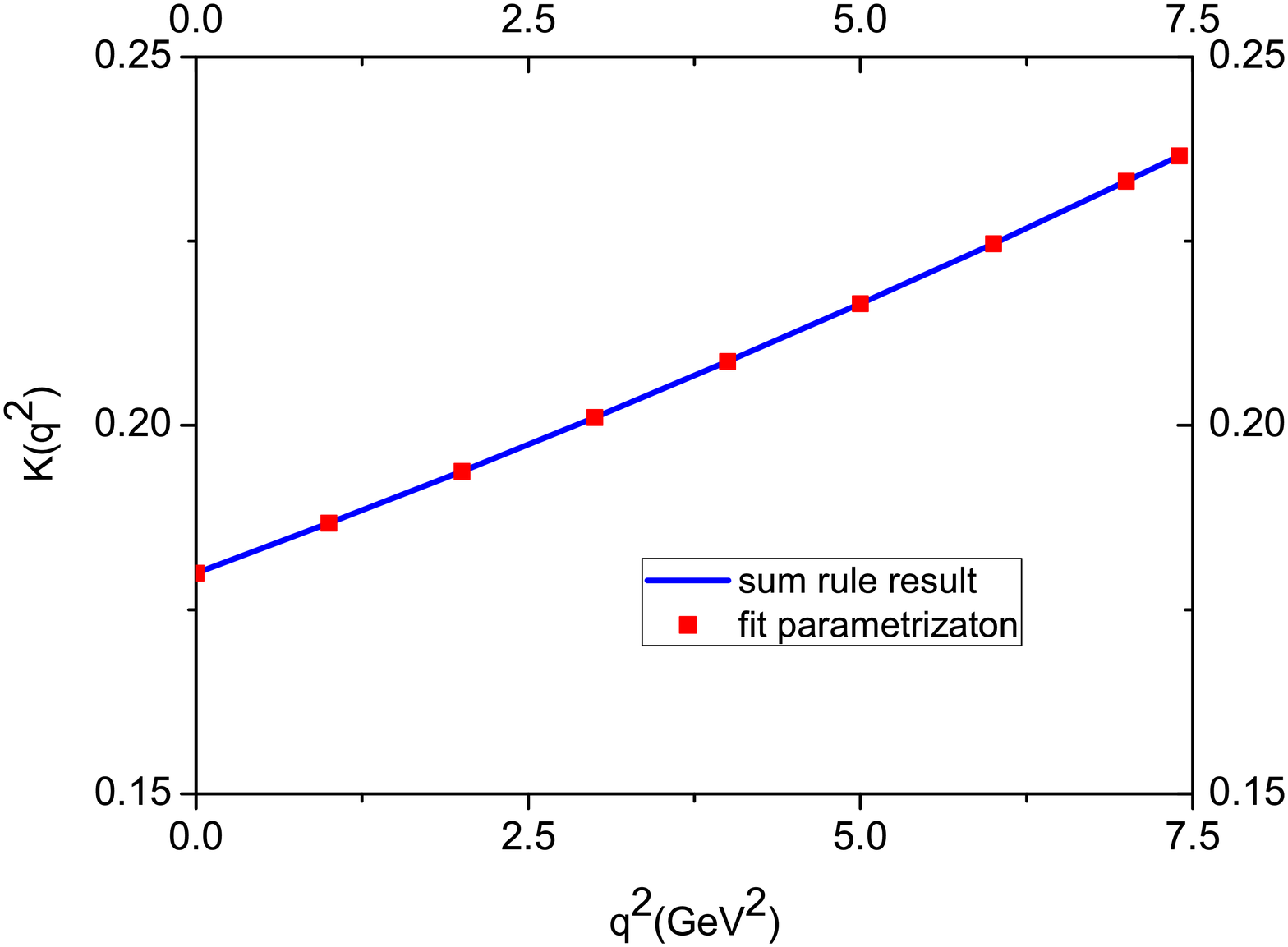}
\end{center}
\caption{
 K($q^2$) as a function of $q^2$ at
$M^2=20~GeV^2$ and $M^{\prime^2}=10~GeV^2$. }
  \label{Kqsq2}
\end{figure}

Having determined the working regions for the auxiliary parameters we proceed to find the behaviors of the form factors in terms of $q^2$.
Our analysis shows that the form factors are well fitted to the function
\begin{eqnarray}\label{fitfunc}
f(q^2)=f_0
\exp[c_1\frac{q^2}{m_{fit}^2}+c_2\Big(\frac{q^2}{m_{fit}^2}\Big)^2
]
\end{eqnarray}
where the values of the parameters, $f_0$, $c_1$, $c_2$
and $m_{fit}^2$ obtained using $M^2=20~\mbox{GeV$^2$}$ and
$M'^2=10~\mbox{GeV$^2$}$ for $B_c\rightarrow
\chi_{c2} \ell\overline{\nu}$ transition are presented in the Table
\ref{fitparam}.
\begin{table}[h]
\renewcommand{\arraystretch}{1.5}
\addtolength{\arraycolsep}{3pt}
$$
\begin{array}{|c|c|c|c|c|}
\hline \hline
         &f_0 & c_1 & c_2 & m_{fit}^2   \\
\hline
  \mbox{$K (q^2)$} &0.18&1.46&-0.11&39.40 \\
  \hline
  \mbox{$b_{-} (q^2)$} &-0.038~GeV^{-2}&1.92&70.85&39.40 \\
  \hline
  \mbox{$b_{+} (q^2)$} &-0.056~GeV^{-2}&1.01&70.67&39.40 \\
  \hline
  \mbox{$h (q^2)$} &-1.70\times10^{-4}~GeV^{-2}&1.69&1.17&39.40 \\
                    \hline \hline
\end{array}
$$
\caption{Parameters appearing in the fit function of the form
factors.} \label{fitparam}
\renewcommand{\arraystretch}{1}
\addtolength{\arraycolsep}{-1.0pt}
\end{table}
As an example we depict the dependence of the form factor $K(q^2)$
on $q^2$ at $M^2=20~\mbox{GeV$^2$}$ and $M'^2=10~\mbox{GeV$^2$}$
 in Fig.~\ref{Kqsq2}, which shows a good fitting of the sum rules results to those obtained from the above fit function.

Our final  purpose in this section is to obtain the decay width and the branching ratio of the
$ B_c\rightarrow \chi_{c2}\ell\overline{\nu}$ transition.  The differential decay width
for this transition is
obtained as \cite{Wang2}
\begin{eqnarray}\label{decaywidth}
&&\frac{d\Gamma}{dq^2}=\nonumber\\&&\frac{\lambda(m_{B_c}^2,m_{\chi_{c2}}^2,q^2)}{4m_{\chi_{c2}}^2}
\Big(\frac{q^2-m_{\ell}^2}{q^2}\Big)^2\frac{\sqrt{\lambda(m_{B_c}^2,m_{\chi_{c2}}^2,q^2)}G_F^2
V_{cb}^2}{384m_{B_c}^3\pi^3}\Bigg\{\frac{1}{2q^2}\Bigg[3m_{\ell}^2\lambda(m_{B_c}^2,m_{\chi_{c2}}^2,q^2)
[V_0(q^2)]^2
\nonumber \\
&+&(m_{\ell}^2+2q^2)\Bigg|\frac{1}{2m_{\chi_{c2}}}\Big[(m_{B_c}^2-m_{\chi_{c2}}^2-q^2)(m_{B_c}-m_{\chi_{c2}})V_1(q^2)
-\frac{\lambda(m_{B_c}^2,m_{\chi_{c2}}^2,q^2)}{m_{B_c}-m_{\chi_{c2}}}V_2(q^2)\Big]\Bigg|^2\Bigg]
\nonumber \\
&+&\frac{2}{3}(m_{\ell}^2+2q^2)\lambda(m_{B_c}^2,m_{\chi_{c2}}^2,q^2)\Bigg[\Bigg|
\frac{A(q^2)}{m_{B_c}-m_{\chi_{c2}}}-\frac{(m_{B_c}-m_{\chi_{c2}})V_1(q^2)}{\sqrt{\lambda(m_{B_c}^2,m_{\chi_{c2}}^2,q^2)}}
\Big|^2
\nonumber \\
&+&\Big|
\frac{A(q^2)}{m_{B_c}-m_{\chi_{c2}}}+\frac{(m_{B_c}-m_{\chi_{c2}})V_1(q^2)}{\sqrt{\lambda(m_{B_c}^2,m_{\chi_{c2}}^2,q^2)}}
\Bigg|^2\Bigg]\Bigg\},
\end{eqnarray}
where
\begin{eqnarray}\label{decaywidth1}
A(q^2)&=&-(m_{B_c}-m_{\chi_{c2}})h(q^2),
\nonumber \\
V_1(q^2)&=&-\frac{k(q^2)}{m_{B_c}-m_{\chi_{c2}}},
\nonumber \\
V_2(q^2)&=&(m_{B_c}-m_{\chi_{c2}})b_{+}(q^2),
\nonumber \\
V_0(q^2)&=&\frac{m_{B_c}-m_{\chi_{c2}}}{2m_{\chi_{c2}}}V_1(q^2)-\frac{m_{B_c}+m_{\chi_{c2}}}{2m_{\chi_{c2}}}V_2(q^2)
-\frac{q^2}{2m_{\chi_{c2}}}b_{-}(q^2).
\end{eqnarray}
After performing integration over $q^2$ in Eq. (\ref{decaywidth}) in
the interval $m_{\ell}^2\leq q^2 \leq (m_{B_c}-m_{\chi_{c2}})^2$, we
obtain the decay widths and the branching ratios as presented in
Table~\ref{numresult} and Table~\ref{table3} for different lepton channels.
We also depict the existing predictions of other non-perturbative approaches on the branching ratio in  Table~\ref{table3}. The results of decay width and branching ratio at
 electron channel are very close to those of the $\mu$ channel so we only present the results at $\mu$ channel. From Table~\ref{table3} we read that our result on the branching ratio  at $\mu$
channel is exactly the same as the prediction of the NRCQM \cite{EHernandez} but it is a bit smaller compared to the CLFQM \cite{Wang2}, GIA \cite{CHChang} and RCQM \cite{MAIvanov,MAIvanov1}.
In the case of $\tau$ channel, our result is comparable with those of GIA \cite{CHChang} and RCQM \cite{MAIvanov1} but it is considerably greater than those of CLFQM \cite{Wang2},  RCQM \cite{MAIvanov}
and NRCQM \cite{EHernandez}.
\begin{table}[h]
\renewcommand{\arraystretch}{1.5}
\addtolength{\arraycolsep}{3pt}
$$
\begin{array}{|c|c|c|}
\hline \hline
         &\Gamma(GeV)  \\
\hline
  \mbox{$B_c \rightarrow \chi_{c2}\tau\overline{\nu}_{\tau}$} &(3.24\pm1.12)\times 10^{-16} \\
  \hline
  \mbox{$B_c \rightarrow \chi_{c2}\mu\overline{\nu}_{\mu}$} &(1.89\pm0.68)\times 10^{-15}\\
                      \hline \hline
\end{array}
$$
\caption{Numerical results of decay width  for different lepton
channels.} \label{numresult}
\renewcommand{\arraystretch}{1}
\addtolength{\arraycolsep}{-1.0pt}
\end{table}
\begin{table}[h]
\renewcommand{\arraystretch}{1.5}
\addtolength{\arraycolsep}{3pt}
$$
\begin{array}{|c|c|c|}
\hline \hline
         &B_c \rightarrow \chi_{c2}\tau\overline{\nu}_{\tau} &
         B_c \rightarrow \chi_{c2}\mu\overline{\nu} \\
\hline
  \mbox{This Work} &(0.020\pm0.007)\times 10^{-2}&(0.130\pm0.048)\times 10^{-2} \\
  \hline
  \mbox{CLFQM\cite{Wang2}} &0.0096^{+0.0047}_{-0.0058}\times 10^{-2}&0.17^{+0.09}_{-0.11}\times 10^{-2} \\
  \hline
  \mbox{GIA\cite{CHChang}} &0.029\times 10^{-2}&0.19\times 10^{-2} \\
   \hline
  \mbox{RCQM\cite{MAIvanov}} &0.0082\times 10^{-2}&0.17\times 10^{-2} \\
  \hline
  \mbox{RCQM\cite{MAIvanov1}} &0.014\times 10^{-2}&0.20\times 10^{-2} \\
  \hline
  \mbox{NRCQM\cite{EHernandez}} &0.0093\times 10^{-2}&0.13\times 10^{-2} \\
                      \hline \hline
\end{array}
$$
\caption{Numerical results of  branching ratio for different lepton
channels.} \label{table3}
\renewcommand{\arraystretch}{1}
\addtolength{\arraycolsep}{-1.0pt}
\end{table}

In summary  we have calculated the form factors responsible for   the transition of  $B_c\rightarrow
\chi_{c2} l \overline{\nu} $ via QCD sum rules in the present  work. We took into account the two-gluon condensate contributions for the   first time in the tensor channel. We also used the fit functions
 of the form factors in terms of $q^2$ to estimate  the decay widths and branching ratios of the
considered transition at different lepton channels. We compared our results of the branching
ratios at different lepton channels  with those of the existing predictions via different non-perturbative approaches. Our results are over all consistent with the predictions of those non-perturbative 
approaches in order of magnitudes. We expect that with the collected or future data at LHC, we will be able to study this decay channel in experiment in near future.

\newpage

\section*{Appendix A}
In this appendix we briefly show how we perform   integrals 
encountered  in the calculations. The types of the integrals are:
\begin{eqnarray}\label{Is}
I_0(a,b,c)=\int\frac{d^4k}{(2\pi)^4}\frac{1}{[k^2-m_1^2]^a[(k+p)^2-m_2^2]^b[(k+p^{\prime})^2-m_3^2]^c},
\nonumber \\
I_{\mu}(a,b,c)=\int\frac{d^4k}{(2\pi)^4}\frac{k_{\mu}}{[k^2-m_1^2]^a[(k+p)^2-m_2^2]^b[(k+p^{\prime})^2-m_3^2]^c},
\nonumber \\
I_{\mu\nu}(a,b,c)=\int\frac{d^4k}{(2\pi)^4}\frac{k_{\mu}k_{\nu}}{[k^2-m_1^2]^a[(k+p)^2-m_2^2]^b[(k+p^{\prime})^2-m_3^2]^c},
\nonumber \\
I_{\mu\nu\alpha}(a,b,c)=\int\frac{d^4k}{(2\pi)^4}\frac{k_{\mu}k_{\nu}k_{\alpha}}{[k^2-m_1^2]^a[(k+p)^2-m_2^2]^b[(k+p^{\prime})^2-m_3^2]^c}.
\end{eqnarray}
To perform these  integrals we use  the Schwinger
representation of  the Euclidean propagator as 
\begin{eqnarray}\label{SchwingerRep}
\frac{1}{(k^2+m^2)^n} = \frac{1}{\Gamma(n)} \int_0^\infty dt
\,t^{n-1} e^{-t(k^2+m^2)}~.
\end{eqnarray}
The four-$k$ integral is performed using the  Gaussian integral and $I_0$
is obtained as
\begin{eqnarray}\label{I0}
I_0(a,b,c)=\frac{i(-1)^{a+b+c}}{16\pi^2\Gamma(a)\Gamma(b)\Gamma(c)}\int_0^\infty
dt_1 t_1^{a-1}\int_0^\infty dt_2 t_2^{b-1}\int_0^\infty dt_3
t_3^{c-1}\frac{e^{-\Delta}}{(t_1+t_2+t_3)^2}
\end{eqnarray}
where
\begin{eqnarray}\label{Delta}
\Delta=\Big(t_2-\frac{t_2^2+t_2t_3}{t_1+t_2+t_3}\Big)p_{E}^2+\Big(t_3-\frac{t_3^2+t_2t_3}{t_1+t_2+t_3}\Big)p_{E}^{\prime
2}+\frac{t_2t_3}{t_1+t_2+t_3}q_{E}^2+t_1m_1^2+t_2m_2^2+t_3m_3^2~.
\nonumber \\
\end{eqnarray}
Application of the Borel transformations with respect to the $p_E^2$
and $p_E^{\prime 2}$ using
\begin{eqnarray}\label{borel}
\hat{{\cal B}}_{p_E^2} (M^2) e^{-\beta p_E^2} = \delta
(1/M^2-\beta)~,
\end{eqnarray}
produces two Dirac Delta functions which help us to perform the
integrals over $t_2$ and $t_3$. Finally, $I_0(a,b,c)$ in the Borel scheme  is obtained as
\begin{eqnarray}\label{eq26}
\hat{{\cal B}}I_{0}(a,b,c) &=&\frac{i(-1)^{a+b+c}}{16\pi
^{2}\,\Gamma
(a)\Gamma (b)\Gamma (c)}(M^{2})^{3-a-b}(M^{\prime 2})^{3-a-c}\,\mathcal{U}%
_{0}(a+b+c-4,1-c-b)~, \nonumber \\
\end{eqnarray}
where
\begin{eqnarray}\label{eq28}
{\cal U}_0(A,B) = \int_0^\infty dy (y+M^2+M^{\prime 2})^A y^B
\,exp\left[ -\frac{B_{-1}}{y} - B_0 - B_{1}y \right]~,
\end{eqnarray}
and
\begin{eqnarray}\label{eq29}
B_{-1} &=& \frac{1}{M^2M^{\prime 2}} \left[m_{3}^2M^4+m_2^2
M^{\prime 4} + M^2M^{\prime 2} (m_2^2+m_{3}^2
-q^2) \right] ~, \nonumber \\
B_0 &=& \frac{1}{M^2 M^{\prime 2}} \left[ M^2(m_{1}^2+m_3^2)  +
M^{\prime 2} (m_1^2+m_2^2)
\right] ~, \nonumber \\
B_{1} &=& \frac{m_1^2}{M^2 M^{\prime 2}}~,
\nonumber \\
y&=&-M^2+M^{\prime 2}(-1+M^2 t_1)~.
\end{eqnarray}

Following similar steps,  $I_{\mu}(a,b,c)$, $I_{\mu\nu}(a,b,c)$ and
$I_{\mu\nu\alpha}(a,b,c)$ are obtained in the Borel scheme  as
\begin{eqnarray}\label{eq26}
\hat{{\cal B}}I_{\mu}(a,b,c) &=&\frac{i(-1)^{a+b+c+1}}{16\pi
^{2}\,\Gamma (a)\Gamma (b)\Gamma
(c)}\mathcal{U}_{0}(a+b+c-5,1-c-b)
\nonumber \\
&\times&\Big[(M^{2})^{3-a-b}(M^{\prime
2})^{4-a-c}\,p_{\mu}+(M^{2})^{4-a-b}(M^{\prime 2})^{3-a-c}
\,p_{\mu}^{\prime}\Big], \nonumber \\
\hat{{\cal
B}}I_{\mu\nu}(a,b,c) &=&\frac{i(-1)^{a+b+c}}{16\pi ^{2}\,\Gamma
(a)\Gamma (b)\Gamma (c)}\Big\{\frac{1}{2}(M^{2})^{4-a-b}(M^{\prime
2})^{4-a-c}~\mathcal{U}_{0}(a+b+c-6,2-c-b)~g_{\mu\nu}
\nonumber \\
&+& (M^{2})^{3-a-b}(M^{\prime
2})^{5-a-c}~\mathcal{U}_{0}(a+b+c-6,1-c-b)~p_{\mu}p_{\nu}
\nonumber \\
&+& (M^{2})^{5-a-b}(M^{\prime
2})^{3-a-c}~\mathcal{U}_{0}(a+b+c-6,1-c-b)~p_{\mu}^{\prime}p_{\nu}^{\prime}
\nonumber \\
&+&(M^{2})^{4-a-b}(M^{\prime
2})^{4-a-c}~\mathcal{U}_{0}(a+b+c-6,1-c-b)~p_{\mu}p_{\nu}^{\prime}
\nonumber \\
&+&(M^{2})^{4-a-b}(M^{\prime
2})^{4-a-c}~\mathcal{U}_{0}(a+b+c-6,1-c-b)~p_{\mu}^{\prime}p_{\nu}
\Big\},
\nonumber \\
\hat{{\cal B}}I_{\mu\nu\alpha}(a,b,c)
&=&\frac{i(-1)^{a+b+c+1}}{16\pi ^{2}\,\Gamma (a)\Gamma (b)\Gamma
(c)}\Big\{(M^{2})^{1-a-b}(M^{\prime
2})^{4-a-c}~\mathcal{U}_{0}(a+b+c-7,1-c-b)~p_{\mu}p_{\nu}p_{\alpha}
\nonumber \\
&+& (M^{2})^{2-a-b}(M^{\prime
2})^{3-a-c}~\mathcal{U}_{0}(a+b+c-7,1-c-b)~p_{\mu}p_{\nu}p_{\alpha}^{\prime}
\nonumber \\
&+& (M^{2})^{2-a-b}(M^{\prime
2})^{3-a-c}~\mathcal{U}_{0}(a+b+c-7,1-c-b)~p_{\mu}^{\prime}p_{\nu}p_{\alpha}
\nonumber \\
&+&(M^{2})^{3-a-b}(M^{\prime
2})^{2-a-c}~\mathcal{U}_{0}(a+b+c-7,1-c-b)~p_{\mu}^{\prime}p_{\nu}p_{\alpha}^{\prime}
\nonumber \\
&+&(M^{2})^{2-a-b}(M^{\prime
2})^{3-a-c}~\mathcal{U}_{0}(a+b+c-7,1-c-b)~p_{\mu}p_{\nu}^{\prime}p_{\alpha}
\nonumber \\
&+&(M^{2})^{3-a-b}(M^{\prime2})^{2-a-c}~\mathcal{U}_{0}(a+b+c-7,1-c-b)~p_{\mu}p_{
\nu}^{\prime}p_{\alpha}^{\prime}
\nonumber \\
&+&(M^{2})^{3-a-b}(M^{\prime
2})^{2-a-c}~\mathcal{U}_{0}(a+b+c-7,1-c-b)~p_{ \mu}^{\prime}p_{
\nu}^{\prime}p_{\alpha}
\nonumber \\
&+&(M^{2})^{4-a-b}(M^{\prime
2})^{1-a-c}~\mathcal{U}_{0}(a+b+c-7,1-c-b)~p_{\mu}^{\prime}p_{\nu}^{\prime}p_{\alpha}^{\prime}
\nonumber \\
&+& \frac{1}{2}(M^{2})^{2-a-b}(M^{\prime
2})^{3-a-c}~\mathcal{U}_{0}(a+b+c-7,2-c-b)~p_{\nu}g_{\mu\alpha}
\nonumber \\
&+&\frac{1}{2}(M^{2})^{3-a-b}(M^{\prime
2})^{2-a-c}~\mathcal{U}_{0}(a+b+c-7,2-c-b)~p_{\nu}^{\prime}g_{\mu\alpha}
\nonumber \\
&+&\frac{1}{2}(M^{2})^{2-a-b}(M^{\prime
2})^{3-a-c}~\mathcal{U}_{0}(a+b+c-7,2-c-b)~p_{\mu}g_{\nu\alpha}
\nonumber \\
&+& \frac{1}{2}(M^{2})^{3-a-b}(M^{\prime
2})^{2-a-c}~\mathcal{U}_{0}(a+b+c-7,2-c-b)~p_{\mu}^{\prime}g_{\nu\alpha}
\nonumber \\
&+&  \frac{1}{2}(M^{2})^{2-a-b}(M^{\prime
2})^{3-a-c}~\mathcal{U}_{0}(a+b+c-7,2-c-b)~p_{\alpha}g_{\mu\nu}
\nonumber \\
&+&   \frac{1}{2}(M^{2})^{3-a-b}(M^{\prime
2})^{2-a-c}~\mathcal{U}_{0}(a+b+c-7,2-c-b)~p_{\alpha}^{\prime}g_{\mu\nu}
\Big\}.
\end{eqnarray}
\section*{Appendix B}
In this appendix we present the explicit expression for the function $\widehat{\textbf{B}}\Pi_1^{non-pert}$, which is given by
\begin{eqnarray}\label{nonpert}
&&\widehat{\textbf{B}}\Pi_1^{non-pert}=\nnb\\&&-\int^{\infty}_{0}dt~~\langle
\frac{\alpha_s G^2}{\pi}\rangle
\exp\Big[f(M^2,M^{\prime^2},t)\Big]\Bigg\{\frac{(M^2-M^{\prime^2})}{48t^2(M^2+M^{\prime^2}+t)^2}
\Big[2tM^{\prime^2}(m_b-2m_c)
\nonumber \\
&-& t^2m_c-m_bM^{\prime^4}+M^2(m_b-3m_c) (M^{\prime^2}+t)\Big]
+\frac{1}{192t^3M^2M^{\prime^2}(M^2+M^{\prime^2}+t)}
\nonumber \\
&\times& \Big[M^8\Big(3M^{\prime^2}(m_b-2m_c)
+m_c(4m_c^2-4m_bm_c-9t)\Big) +
M^4M^{\prime^2}\Big(17tm_cM^{\prime^2}
\nonumber \\
&-& M^{\prime^4}(m_b-6m_c)
+t^2(21m_c-5m_b)\Big)+M^2\Big(6m_bM^{\prime^8}+3t^2M^{\prime^4}(7m_b-6m_c)
\nonumber \\
&+& 4t^3m_c^2(m_c-m_b)
+M^{\prime^6}(4m_cm_b^2-4m_b^3+15tm_b-18tm_c)\Big)-M^{\prime^6}\Big(6m_b
M^{\prime^4}
\nonumber \\
&+&
9t^2m_c+M^{\prime^2}(4m_c^3-4m_bm_c^2+3tm_b-10tm_c)\Big)-M^{\prime^2}
\Big(2m_bM^{\prime^8}+3t^2M^{\prime^4}
\nonumber \\
&\times& (3m_b+m_c)+4t^3m_c^2(m_c-m_b)+M^{\prime^6}(4m_b^2m_c
-4m_b^3+11tm_b+3tm_c)\Big)\Big]
\nonumber \\
&-&\frac{tm_c^3}{384M^4M^{\prime^4}}\Big(m_b^2+2m_bm_c+5m_c^2-q^2\Big)
-\frac{m_c^3}{192tM^2M^{\prime^2}}\Big(m_b^2+4m_bm_c+m_c^2-q^2\Big)
\nonumber \\
&-&\frac{m_b^2M^{\prime^4}}{384t^3M^4}\Big(13m_bM^{\prime^2}+m_c(5m_b^2+10m_bm_c+7m_c^2-4q^2)\Big)
+\frac{1}{384t^2M^2}
\nonumber \\
&\times&\Big[M^{\prime^4}(23m_b+6m_c)-m_bm_c^2(m_b^2+2m_bm_c+3m_c^2)+M^{\prime^2}(3m_b^3+18m_cm_b^2
\nonumber \\
&+&
31m_bm_c^2+11m_c^3-2m_bq^2-11m_cq^2)\Big]+\frac{m_c}{384t^2M^{\prime^2}}\Big[21M^4
+m_c^2(m_b^2-6m_bm_c
\nonumber \\
&+&
3m_c^2)+M^2(3m_b^2+5m_bm_c+7m_c^2-3q^2)\Big]-\frac{m_b^2M^{\prime^2}}{384t^2M^2}\Big[2M^{\prime^2}
(m_b-m_c)
\nonumber \\
&+&m_c(m_b^2+8m_c^2-q^2)\Big]+\frac{m_c^3M^2}{384t^2M^{\prime^2}}\Big(2M^2-m_b^2-m_bm_c
-3m_c^2+q^2\Big)
\nonumber \\
&-&\frac{m_b^2M^{\prime^6}}{384t^4M^4}\Big[9m_bM^{\prime^2}+m_c(7m_b^2+6m_bm_c+3m_c^2-3q^2)\Big]+
\frac{m_c^3M^6}{384t^4M^{\prime^4}}\Big(3M^2+m_b^2
\nonumber \\
&+&2m_bm_c+5m_c^2-q^2\Big)+\frac{m_c^2}{384tM^4}\Big[5m_bM^{\prime^2}-m_c(8m_b^2+6m_bm_c+3m_c^2-q^2)
\Big]
\nonumber \\
&+&\frac{m_c^3}{384tM^{\prime^4}}\Big(3M^2+m_b^2+m_bm_c+5m_c^2-q^2M2sq^3
t^5\Big)+\frac{m_bM^{\prime^2}}{384t^3M^2}\Big[4M^{\prime^4}-2m_bm_c^3
\nonumber \\
&+&M^{\prime^2}(9m_b^2+27m_bm_c
+4m_c^2-4q^2)\Big]+\frac{m_c^3M^2}{384t^3M^{\prime^2}}\Big[M^2(17m_c-3m_b)-m_c(5m_b^2
\nonumber \\
&+& 10m_bm_c+7m_c^2
-4q^2)\Big]+\frac{m_c^2}{384M^4M^{\prime^2}}\Big[4m_cq^2-5m_b^2m_c-7m_c^3+2m_b(t-5m_c^2)\Big]
\nonumber \\
&+&\frac{tm_c^3-m_c^5}{192M^2M^{\prime^4}}+\frac{m_b^2M^{\prime^4}}{384t^4M^2}\Big[3M^{\prime^2}
(m_b+2m_c)-2m_c(m_b^2+4m_bm_c+m_c^2-q^2)\Big]
\nonumber \\
&-&\frac{m_c^2M^4}{384t^4M^{\prime^2}}\Big[M^2(2m_b+9m_c)
+2m_c(m_b^2+4m_bm_c+m_c^2-q^2)\Big]+\frac{13m_c^3-5m_bm_c^2}{384M^2M^{\prime^2}}
\nonumber \\
&-&\frac{m_b^3M^{\prime^6}}{384t^5M^2}\Big(m_b^2+2m_bm_c+3m_c^2\Big)+\frac{m_c^3M^6}{384t^5M^{\prime^2}}
\Big(m_b^2-2m_bm_c+3m_c^2\Big)+\frac{m_c}{384tM^2}
\nonumber \\
&\times&
\Big(3m_b^2+5m_bm_c+12m_c^2-3q^2\Big)-\frac{m_c}{384tM^{\prime^2}}\Big(3m_b^2+5m_bm_c+18m_c^2-3q^2\Big)
\nonumber \\
&+&\frac{5m_bm_c^2}{384M^4}+\frac{m_b^3m_c^2}{384M^6}-\frac{m_c^3}{192M^4}-\frac{m_c^3}{128M^{\prime^4}}
-\frac{m_bm_c^4}{128M^6}+\frac{m_c^5}{192M^{\prime^6}}+\frac{m_b^5M^{\prime^8}}{192t^5M^4}
\nonumber \\
&+&\frac{m_b^5M^{\prime^{10}}}{128t^5M^6} -
\frac{m_cm_b^4M^{\prime^8}}{128t^5M^4}
+\frac{m_bm_c^4M^8}{384t^5M^{\prime^4}}-\frac{m_c^5M^{10}}{384t^5M^{\prime^6}}+\frac{m_c^5M^8}{192t^5M^{\prime^4}}
+\frac{m_b^5M^{\prime^8}}{96t^4M^6}
\nonumber \\
&+& \frac{m_b^5M^{\prime^6}}{384t^3M^6}
-\frac{m_b^3m_c^2M^{\prime^6}}{128t^3M^6}
-\frac{m_c^3M^6}{128t^3M^{\prime^4}}+\frac{m_c^5M^6}{192t^3M^{\prime^6}}
-\frac{m_c^5M^4}{192t^3M^{\prime^4}}-\frac{m_b^3m_c^2M^{\prime^4}}{384t^2M^6}
\nonumber \\
&-& \frac{m_c^5M^4}{384t^2M^{\prime^6}}
+\frac{m_b^3m_c^2M^{\prime^2}}{128tM^6}-\frac{m_c^5M^2}{384tM^{\prime^6}}
-\frac{tm_bm_c^4}{96M^6M^{\prime^2}}
-\frac{t^2m_bm_c^4}{384M^6M^{\prime^4}}-\frac{t^2m_c^5}{384M^4M^{\prime^6}}
\nonumber \\
&+&\frac{m_bm_c}{384t^5}\Big[M^2M^{\prime^2}m_bm_c(3m_b-m_c)+M^{\prime^4}m_b(m_b^2+m_cm_b+3m_c^2)
-M^4m_c(m_b^2
\nonumber \\
&+&
3m_bm_c+3m_c^2)\Big]+\frac{m_c}{384t^4}\Big[3m_bm_cM^4-m_b^2M^{\prime^2}(M^{\prime^2}-
m_b^2-6m_bm_c-5m_c^2+q^2)\Big]
\nonumber \\
&-& M^2\Big(m_bM^{\prime^2}(7m_c-5m_b)+m_c^2(7m_b^2+6m_bm_c
+3m_c^2-3q^2)\Big)-\frac{1}{384t^3}\Big[23M^{\prime^4}m_b
\nonumber \\
&-&
M^4(17m_c-6m_b)-m_bm_c^2(m_b^2+2m_bm_c-3m_c^2)+M^{\prime^2}(5m_b^3+12m_b^2m_c+7m_bm_c^2
\nonumber \\
&+& 3m_c^3
-4m_bq^2-3m_cq^2)+M^2m_c(8m_b^2+3m_bm_c-10m_c^2-5q^2)-18M^2M^{\prime^2}m_b
\nonumber \\
&+&12M^2M^{\prime^2}m_c\Big]+\frac{1}{384t^2}\Big[9M^2(m_b-6m_c)-34M^{\prime^2}m_b-2m_b^3
+19M^{\prime^2}m_c-16m_b^2m_c
\nonumber \\
&-& 24m_bm_c^2-26m_c^3+2m_bq^2+13m_cq^2\Big]+\frac{m_b-m_c}{384t}
 \Bigg\},
\end{eqnarray}
where
\begin{eqnarray}\label{L}
f(M^2,M^{\prime^2},t)&=&-\frac{2M^2m_c^2+M^{\prime^2}(m_b^2+m_c^2)}{M^2M^{\prime^2}}
-\frac{M^{\prime^4}m_b^2+M^4m_c^2+M^2M^{\prime^2}(m_b^2+m_c^2-q^2)}{M^2M^{\prime^2}t}
\nonumber \\
&-& \frac{m_c^2~~t}{M^2M^{\prime^2}}.
\end{eqnarray}


\newpage

\begin{thebibliography}{99}

\bibitem{Abe} F. Abe et al, Phys. Rev. D58 (1998) 112004; F. Abe et al, Phys. Rev. Lett. 81 (1998) 2432.
\bibitem{Aaltonen} T. Aaltonen et al, Phys. Rev. Lett. 100 (2008) 182002.
\bibitem{Abazov} V. M. Abazov et al, Phys. Rev. Lett. 101 (2008) 012001.
\bibitem{JBeringer} J. Beringer et al. (Particle Data Group), Phys. Rev. D86, 010001 (2012).

\bibitem{AlievKazim} T. M. Aliev, K. Azizi, M. Savci, Phys. Lett. B 690, (2010) 164-167.
\bibitem{Wang2} X. Wang, W. Wang, Phys. Rev. D 79, 114018, (2009).
\bibitem{CHChang}C. H. Chang, Y. Q. Chen, G. L. Wang and H. S. Zong, Phys. Rev. D 65, 014017, (2002).
\bibitem{MAIvanov}M. A. Ivanov, J. G. Korner and P. Santorelli, Phys. Rev. D 73, 054024, (2006).
\bibitem{MAIvanov1}M. A. Ivanov, J. G. Korner and P. Santorelli, Phys. Rev. D 71, 094006, (2005)
[Erratum-ibid. D 75, 019901 (2007)].
\bibitem{EHernandez}E. Hernandez, J. Nieves and J. M. Verde-Velasco, Phys. Rev. D 74, 074008, (2006).


\bibitem{Aliev1} T. M. Aliev, M. Savci, Phys. Lett. B 434 (1998) 358.
\bibitem{Aliev4} T. M. Aliev, M. Savci, Phys. Lett. B 480 (2000) 97.
\bibitem{Azizi2} K. Azizi, R. Khosravi Phys. Rev. D 78 036005 (2008).
\bibitem{Azizi3} K. Azizi, F. Falahati, V. Bashiry, S. M. Zebarjad Phys. Rev. D 77 114024 (2008).
\bibitem{Azizi5} K. Azizi, V. Bashiry, Phys. Rev. D 76 114007 (2007).
\bibitem{Azizi6} K. Azizi, H. Sundu, M. Bayar, Phys. Rev D 79, 116001 (2009).
\bibitem{Colangelo} P. Colangelo, G. Nardulli, and N. Paver, Z. Phys. C 57,
43 (1993).
\bibitem{Kiselev} V. V. Kiselev, A. E. Kovalsky, and A. K. Likhoded, Nucl.
Phys. B 585, 353 (2000); V. V. Kiselev, A. K. Likhoded,
and A. I. Onishchenko, Nucl. Phys. B 569, 473 (2000).
\bibitem{Huang} T. Huang and F. Zuo, Eur.Phys.J.C 51, 833 (2007).

\bibitem{Ivanov} M. A. Ivanov, J. G. Korner and P. Santorelli, Phys. Rev. D 73 054024 (2006).
\bibitem{Ebert1} D. Ebert, R.N. Faustov, V.O. Galkin, Phys. Rev. D 68, 094020 (2003).
\bibitem{Ebert2} D. Ebert, R.N. Faustov, V.O. Galkin, Eur. Phys. J. C 32, 29 (2003).
\bibitem{Faessler} A. Faessler, Th. Gutsche, M. A. Ivanov, J. G. Korner, V. E. Lyubovitskij, Eur. Phys. J.direct C 4, 18(2002).

\bibitem{Chang} C. H. Chang, Y. Q. Chen, G. L. Wang and H. S. Zong, Phys. Rev. D 65, 014017 (2002).
\bibitem{Ivanov2} M. A. Ivanov, J. G. Korner and P. Santorelli, Phys. Rev. D 71, 094006 (2005) [Erratum-ibid. D 75, 019901 (2007)].
\bibitem{Hernandez} E. Hernandez, J. Nieves and J. M. Verde-Velasco, Phys. Rev. D 74, 074008 (2006) [arXiv:hep-ph/0607150].

\bibitem{Becher} T. Becher, H. Boos, E. Lunghi, JHEP0712:062 (2007).
\bibitem{Aglietti} U. Aglietti, L. D. Giustino, G. Ferrera, A. Renzaglia, G. Ricciardi, L. Trentadue, Phys. Lett. B 653 38-52 (2007).
\bibitem{Schwanda} C. Schwanda, Belle Collaboration, Phys. Rev. D 75 032005 (2007).
\bibitem{Chang2} Chao-Hsi Chang, Yu-Qi Chen, Guo-Li Wang, Hong-Shi Zong, Phys. Rev. D 65 014017 (2002).
\bibitem{Colangelo2} P. Colangelo, F. De Fazio, T.N. Pham, Phys. Rev. D 69 054023 (2004).

\bibitem{Reinders85} L. J. Reinders, H.
Rubinstein and S. Yazaki, Phys. Rept. {\bf 127} (1985) 1.

\bibitem{Veliev} E. Veli Veliev, K. Azizi, H. Sundu, N. Aksit, J. Phys. G39, (2012) 015002.





\end{thebibliography}
\end{document}